\documentclass[sigconf]{acmart}

\newcommand\blfootnote[1]{%
	\begingroup
	\renewcommand\thefootnote{}\footnote{#1}%
	\addtocounter{footnote}{-1}%
	\endgroup
}

\setcopyright{none}
\renewcommand\footnotetextcopyrightpermission[1]{} % removes footnote with conference information in first column
\acmConference[IIR 2019]{10th Italian Information Retrieval Workshop}{September 16-18, 2019}{Padova, Italy}
\begin{document}

%%
%% The "title" command has an optional parameter,
%% allowing the author to define a "short title" to be used in page headers.
\title{A Semi-Automated Approach for Information Extraction, \\ Classification and Analysis of Unstructured Data}

%%
%% The "author" command and its associated commands are used to define
%% the authors and their affiliations.
%% Of note is the shared affiliation of the first two authors, and the
%% "authornote" and "authornotemark" commands
%% used to denote shared contribution to the research.
\author{Alberto Purpura}
\email{purpuraa@dei.unipd.it}
\orcid{0000-0003-1701-7805}
\affiliation{%
  \institution{University of Padua}
  \city{Padua}
  \state{Italy}
}

\author{Marco Calaresu}
\email{mcalaresu@uniss.it}
\affiliation{%
	\institution{University of Sassari}
	\city{Sassari}
	\state{Italy}
}

%%
%% By default, the full list of authors will be used in the page
%% headers. Often, this list is too long, and will overlap
%% other information printed in the page headers. This command allows
%% the author to define a more concise list
%% of authors' names for this purpose.
%\renewcommand{\shortauthors}{Trovato and Tobin, et al.}

%%
%% The abstract is a short summary of the work to be presented in the
%% article.
\begin{abstract}
	In this paper, we show how Quantitative Narrative Analysis \cite{franzosi2010quantitative} and simple Natural Language Processing techniques apply to the extraction and categorization of data in a sample case study of the Diary of the former President of the Italian Republic (PoR), Giorgio Napolitano. The Diary contains a record of all his institutional meetings. This information, if properly handled, allows for an analysis of how the PoR used his so-called soft-powers to influence the Italian political system during his first mandate. In this paper, we propose a way to use simple, yet very effective, Natural Language Processing techniques --  such as Regular Expressions and Named Entity Recognition -- to extract information from the Diary. Then, we propose an innovative way to organize the extracted data relying on the methodological framework of Quantitative Narrative Analysis. Finally, we show how to analyze the structured data under different levels of detail using PC-ACE (Program for Computer-Assisted Coding of Events), a software developed specifically for this task and for data visualization. 	\blfootnote{This work was supported by the CDC-STARS project and co-funded by UNIPD.} 
\end{abstract}

%%
%% The code below is generated by the tool at http://dl.acm.org/ccs.cfm.
%% Please copy and paste the code instead of the example below.
%%
\begin{CCSXML}
	<ccs2012>
	<concept>
	<concept_id>10002950.10003648.10003688.10003699</concept_id>
	<concept_desc>Mathematics of computing~Exploratory data analysis</concept_desc>
	<concept_significance>500</concept_significance>
	</concept>
	<concept>
	<concept_id>10010147.10010178.10010179.10003352</concept_id>
	<concept_desc>Computing methodologies~Information extraction</concept_desc>
	<concept_significance>500</concept_significance>
	</concept>
	<concept>
	<concept_id>10002951.10003317.10003347.10003356</concept_id>
	<concept_desc>Information systems~Clustering and classification</concept_desc>
	<concept_significance>300</concept_significance>
	</concept>
	</ccs2012>
\end{CCSXML}

\ccsdesc[500]{Mathematics of computing~Exploratory data analysis}
\ccsdesc[500]{Computing methodologies~Information extraction}
\ccsdesc[300]{Information systems~Clustering and classification}
%%
%% Keywords. The author(s) should pick words that accurately describe
%% the work being presented. Separate the keywords with commas.
\keywords{Natural Language Processing, Data Analysis, Information Extraction}

%%
%% This command processes the author and affiliation and title
%% information and builds the first part of the formatted document.
\maketitle

\section{Introduction}

Researchers in the human sciences, such as political or social scientists, often rely on information contained in textual documents such as newspaper articles or other narrative texts for their studies. However, few of them have been analyzing these sources in a systematical way using automated or semi-automated content analysis techniques, even if this has been proven to be a fruitful method \cite{hopkins2010method, tebaldi2018chi}.
In this research context, Quantitative Narrative Analysis (QNA) is a technique which acts as a bridge between the social sciences and data science. Originally developed by Roberto Franzosi \cite{franzosi2010quantitative}, this technique turns ``words to numbers''. In other terms, it grants a flexible structure for the analysis of complex systems (such as the power relations in the sample case study we describe in Section \ref{sec:scs}) or historical events \cite{franzosi1998narrative} (such as the rise of Fascism in Italy \cite{franzosi1997mobilization}). 
QNA has relied traditionally on the well known five Ws + H (Who, What, Where, When, Why and How) of journalism and translated them into an arbitrarily complex user-defined hierarchical structure which can be stored in a relational database \cite{franzosi2010quantitative}.
It was developed with the goal of analyzing complex events -- considering all available narrative sources, among which the easiest to obtain is usually newspaper articles -- involving a large number of actors, each carrying its own information. For instance, if we analyzed a series of strikes \cite{franzosi1989one}, an actor involved in such events might be a worker or the owner of a factory, each possibly with a different political orientation and background which might be important to consider when evaluating his/her actions.\\
This study proposes a novel way to integrate QNA with Natural Language Processing (NLP) techniques to work in synergy to simplify -- and in the future also to automate -- the analysis of complex systems, in particular in the field of political and social sciences. We show how to extend the framework of QNA in order to analyze non-narrative text, opening new research paths in this field. We also demonstrate how this extension of the theoretical framework can be applied to a case study. We consider a sample case study of the Diary of the Italian President of the Republic (PoR) Giorgio Napolitano during his first presidential term. This Diary is a collection of the meetings in the PoR's political agenda, as shown in Table \ref{tab:sample_diary}. This case study serves as a pretext to show how simple NLP techniques can be applied in order to extract information from a textual source which can be later analyzed using QNA. 

The paper is organized as follows. In Section \ref{sec:rw}, we frame this approach in the context of quantitative research in the human sciences, in particular in political and social science. In Section \ref{sec:scs}, we present the chosen case study of the  Diary of the PoR; in Section \ref{sec:ie} we describe the proposed approach for information extraction. In Section \ref{sec:da} we present how the extracted information can be analyzed at different levels of detail and visualized employing PC-ACE. \footnote{PC-ACE is available for free at \url{www.pc-ace.com}.} Finally, in Section \ref{sec:cfc} we conclude and present some challenges worth exploring in the future.

\section{Related Work}
\label{sec:rw}
QNA is a social science research method for the analysis of narrative text \cite{franzosi2010quantitative}. It does so via a coding scheme (a ``story grammar'') made up of coding categories (the elements of the grammar) to which human coders assign specific portions of a text \cite{franzosi2013quantitative}.
By computing the frequencies and correlations between these coding categories, words are turned to numbers which can be analyzed with statistical tools. Contrary to content analysis \cite{elo2008qualitative}, QNA's coding categories and schemes are based on invariant, structural properties of narrative, namely the sequential organization of narrative (in story and plot and the five Ws of journalism) and its simple surface linguistic structure of subject-verb-object (SVO)~\cite{franzosi1998narrative}.

While most of the research works in social or political science such as \cite{tebaldi2014power} rely on quantitative measurements after a qualitative assessment and classification of the data, QNA allows for a more fine-grained analysis which can be adjusted to different levels of detail. However, an analysis of such large quantities of data becomes impossible to do in a qualitative way on certain levels of detail without the help of a system able to (i) contain and process efficiently all of these information, and (ii) at the same time accurately represent the relations between different entities and their characteristics. The usage of QNA for the analysis of large quantities of data in these terms guided the development of specialized software such as PC-ACE, MAXQDA or NVivo \cite{franzosi2013quantitative}. However, the development of new and always more efficient Natural Language Processing (NLP) techniques \cite{manning2014stanford}, opened new possibilities for this type of analysis which traditionally has relied on ``human coders'' for the process of information extraction and data categorization. In this paper, we propose a way to fill the gap between the human sciences and NLP.\\

\section{Sample Case Study}
\label{sec:scs}
In this section, we will give a general background for the interpretation of our case study, the analysis of the Diary of the President of the Republic (PoR).
%In Italy, the PoR is elected by the parliament for a seven-year mandate, two years longer than the Chamber of Deputies (lower house) and the Senate of the Republic (higher house). 

The Italian Constitution assigns the PoR, as Head of the State, a role that is not only symbolic or ceremonial in nature, but also endowed with high political relevance. The PoR’s power in the Italian parliamentary system has been mainly analyzed through three different theoretical approaches: the \emph{institutional}, \emph{relational} and \emph{presidential leadership} approaches. 

The institutional approach analyzes the presidential power as a result of the normative restraints and intervention opportunities conferred to the PoR by the Constitution, and by established institutional practices. %The subject of this theoretical approach is not the power of the PoR itself, but the formal and informal powers that each president may use (or not) in the relations with other relevant political and institutional actors. 

The relational approach analyzes the power as a result of the PoR’s relations with other political actors. %In consideration of the conciseness of the Constitution articles regarding the PoR -- and also due to the high degree of discretion conferred by the Constitution to the PoR on the use of his/her powers -- some authors state that the presidential power depends on his/her relations with other political (or politically relevant) actors. 
Since political parties are considered by several scholars as the most important actors of reference, the variability of the PoR's power ultimately depends on the relations of the PoR with the parties; i.e. the so-called ``presidential accordion''~\cite{tebaldi2005presidente}. 

Recently, an interpretation of the PoR power was proposed which develops within the \textit{leadership theory}. The theory refers to the processes of presidentialization of politics~\cite{poguntke2007presidentialization}, and analyzes the PoR’s power as a result of exogenous conditions (e.g., international factors such as the EU restraints or opportunities) and/or endogenous conditions (e.g., cultural factors such as the disaffection from party politics and governmental institutions) to the political system. These conditions enable the PoR to decide whether or not to use his/her personal resources in power relations. The PoR’s higher power would show through a more intensive use of the so-called informal or soft powers. These soft-powers are based on his/her personal communication skills and personal resources exercised through formal and informal channels of influence, like the freedom of speech and expression of his/her personal opinions, on every possible policy issue~\cite{tebaldi2014notary}, and the ``moral suasion'' powers~\cite{ amoretti2014power, palladino2015presidentialisations}.

To analyze the above factors which describe the presidential powers according to the leadership theory, we choose to employ the theoretical framework offered by QNA. The time period taken into consideration for the analysis corresponds to the first Napolitano’s Presidency, and stretches from May 15 2006 to April 30 2013. Information on 3068 events of the seven-year term -- contained in the Diary~\footnote{The Diary is available online at: \url{http://presidenti.quirinale.it/elementi/Elenchi.aspx?tipo=Visita}.} -- was then collected automatically from the Web. The PoR's Diary contains all of the appointments of the President, written in a formal style in Italian as shown in Table \ref{tab:sample_diary}.  This repetitive and formal style of the writing was exploited to our advantage to extract information from it, as detailed in Section \ref{sec:ie}. However, our information extraction approach could of course be adapted to any other type of textual data.
\begin{table}[htbp]
	\small
	\centering
	\begin{tabular}{|l|p{1.5cm}|p{5cm}|}
		\hline
		Date & Place & Description \\ \hline
		5/30/06 & Palazzo del Quirinale & On. Sen. Franco MARINI, Presidente del Senato della Repubblica, e On. Fausto BERTINOTTI, Presidente della Camera dei Deputati  \\ \hline
		6/7/06  & Palazzo del Quirinale & On. Silvio BERLUSCONI, Presidente di Forza Italia                                                                                                                       \\ \hline
		6/3/08  & Palazzo della FAO     & Intervento alla cerimonia di apertura della Conferenza sulla,sicurezza alimentare, promossa dalla FAO                                                                   \\ \hline
	\end{tabular}
	\caption{Sample records from the Diary of the President of the Italian Republic, Giorgio Napolitano.}
	\label{tab:sample_diary}
	\vspace{-20pt}
\end{table}
Then, we built a hierarchical structure able to contain the information in the Description field of the Diary according to the theory of QNA. For example, the information contained in the second record of Table \ref{tab:sample_diary} can be represented as: \vspace{5pt}\\
\small
(\textbf{Event}: \\ \indent (\textbf{Subject}: \textit{Presidente della Repubblica}), \\ \indent(\textbf{Verb}: \textit{incontra}), 
\\ \indent(\textbf{Object}: \textit{On. Silvio BERLUSCONI}), 
\\\indent(\textbf{Internal Politics}: 
\\ \indent\indent(\textbf{Political Organizations}:
\\\indent \indent\indent(\textbf{Political Parties}: \textit{Leader of party}),\\ \indent \indent\indent
(\textbf{Goverment}: \textit{Prodi II}),
\\\indent\indent \indent(\textbf{Parliamentary/Extraparliamentary}: \textit{Parliamentary}),
\\ \indent\indent \indent(\textbf{Majority/Minority Political Parties}: \textit{Minority}), \\ \indent\indent \indent
(\textbf{Party Name}: \textit{Forza Italia})),
\\ \indent \indent(\textbf{Legislative Power}: 
\\ \indent\indent \indent(\textbf{Chamber of Deputies}: \textit{Leader of Minority Group}))),
\\\indent(\textbf{Date}: 7 Giugno 2008), \\\indent(\textbf{Place}: \textit{Palazzo del Quirinale})).\vspace{5pt} \\
\normalsize
The terms in bold indicate the categories of the hierarchical structure of QNA that we created, and the terms in italics indicate the information which has been extracted directly or indirectly (relying on auxiliary information sources) from the record. This particular type of hierarchical structure, where the categories and the complexity can be arbitrary, represents the different factors involved in the process of interest; and can be used in this case to measure the intensity of the relations of the PoR -- i.e. with the different members of the Italian parliament or with the representatives of the executive and/or legislative power -- by counting the frequency of the meetings between the PoR and the various institutional actors as motivated more in detail in Section \ref{sec:ie}.

\section{Information Extraction and Classification}
\label{sec:ie}
After we represented the key factors of our case study in terms of a hierarchical structure as QNA suggests, we can proceed with the extraction of the information from the source we chose to employ. In this case, we focus on textual documents since this is the most employed source of data for most of the analyzes in the political and social science fields. However, the nature of QNA allows for the extraction of information from any kind of media such as videos or pictures. With the appropriate software for the recognition of entities in images for example, one could easily build his/her own data set in the same way we showed in Section \ref{sec:scs}, i.e. one could choose to represent the entities present in each frame of a video and study the interactions between them with the statistical techniques of data science and under different levels of abstraction/aggregation as we suggest in Section \ref{sec:da}. In this sample case study we opted to show how relevant information for a political science research can be extracted from unstructured text and organized into a hierarchical fashion. 

We developed a  tool~\footnote{\url{https://github.com/albpurpura/AgendaParser}.} in Python, based on Regular Expressions (RE) for Named Entity Recognition (NER) in Italian, -- specifically for the detection of terms identifying persons -- which exploits the style of the Diary to extract the names and roles of each actor the PoR interacts with. For example, we exploit the fact that the surnames of the actors who the PoR meets are all in capitals and that there is a closed vocabulary to indicate the different roles of each person the President interacts with (i.e. ``On.'' for members of the Parliament and ``Presidente'' for leaders of a political party or presidents of an institution such as the Senate). This approach, which relies in this case on the structure of the sentences, could be easily generalized to other styles of writing. For example, using aspect classification techniques if the categories of the entities to extract are known \cite{ws4absa}, or topic modeling approaches \cite{purpura2018nonnegative}.

Next, we employed external sources of information useful for our case study, which we built from data extracted from trusted sources on the Web, to fill the categories in our structure relative for example to the current government or to the party where each each actor belonged to. The structure which holds this information is general enough to represent all the attributes that we are interested to consider in the analysis, for this reason, some attributes can be left empty if they do not apply to the actor we are currently considering (i.e. the PoR is participating to an institutional ceremony as in the third record of Table \ref{tab:sample_diary}). All of this can be attained very easily and in a semi-automated fashion using simple software tools created ad-hoc for the task and a few human verifiers to supervise the process and correct eventual mistakes. Again, this process can be easily adapted to other types of information given an appropriate mapping of the extracted and categorized data to the hierarchical structure which will contain it.

Once we obtained a table where each column represents an entity in our hierarchical structure (i.e. Subject, Verb, etc.) and each row refers to each instance we wish to analyze (i.e. one meeting of the PoR), we can import the data for example in PC-ACE (Program for Computer-Assisted Coding of Events) -- or another specialized software which allows to analyze the data through QNA -- and proceed with a semi-automated analysis of the collected data.

\section{Data Analysis and Visualization with PC-ACE}
\label{sec:da}
\begin{figure}[h!]	
	\vspace{-20pt}
	\includegraphics[scale=0.6]{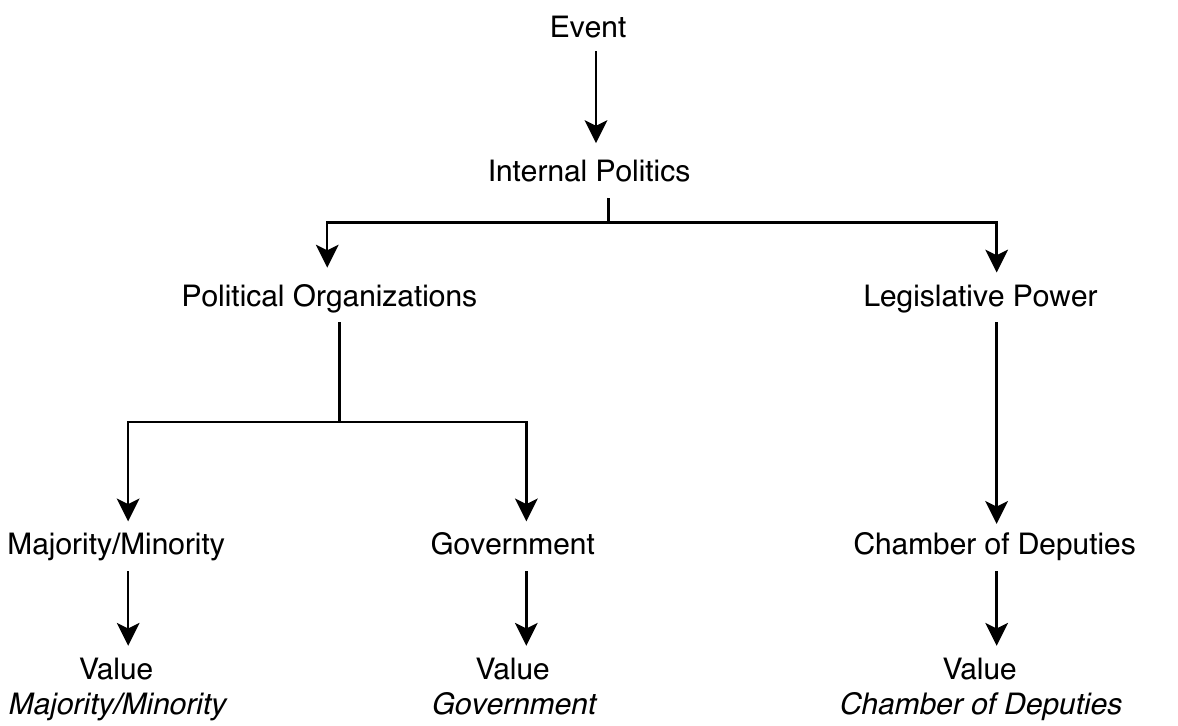}
		\caption{The query manager form in PC-ACE.}
	\label{fig:qmform}
	\vspace{-10pt}
\end{figure}

\begin{figure}[h!]
	\includegraphics[width=9cm]{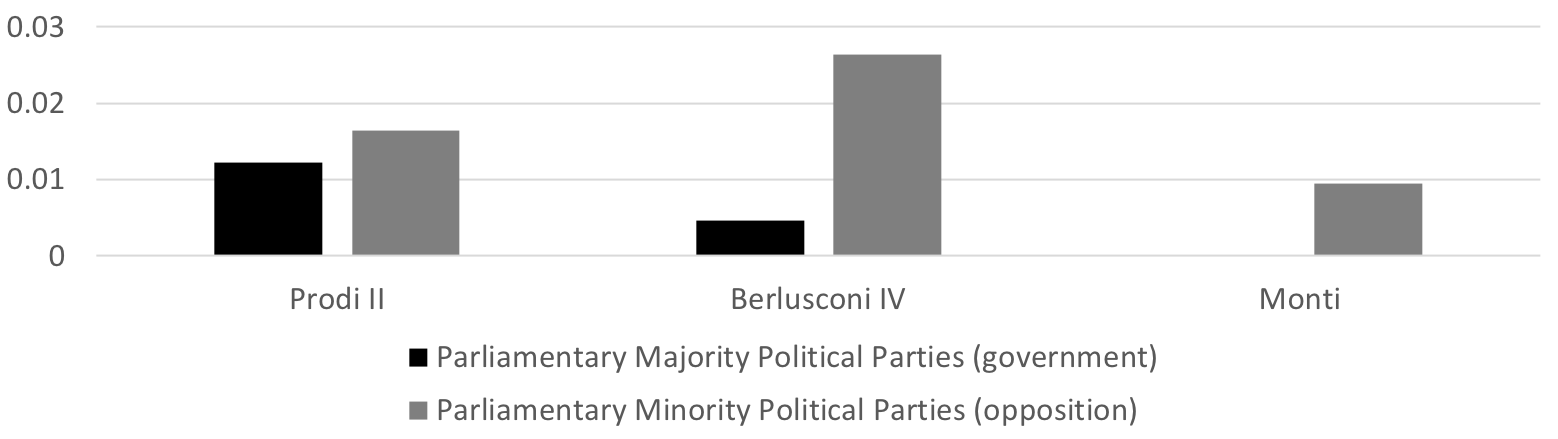}
	\caption{Histogram of the number of meetings with members of the minority and majority parliamentary groups, normalized by government duration.}
	\label{fig:majorityminority}
	\vspace{-10pt}	
\end{figure}

\begin{figure}[htb!]
	\vspace{-5pt}
	\includegraphics[scale=0.38]{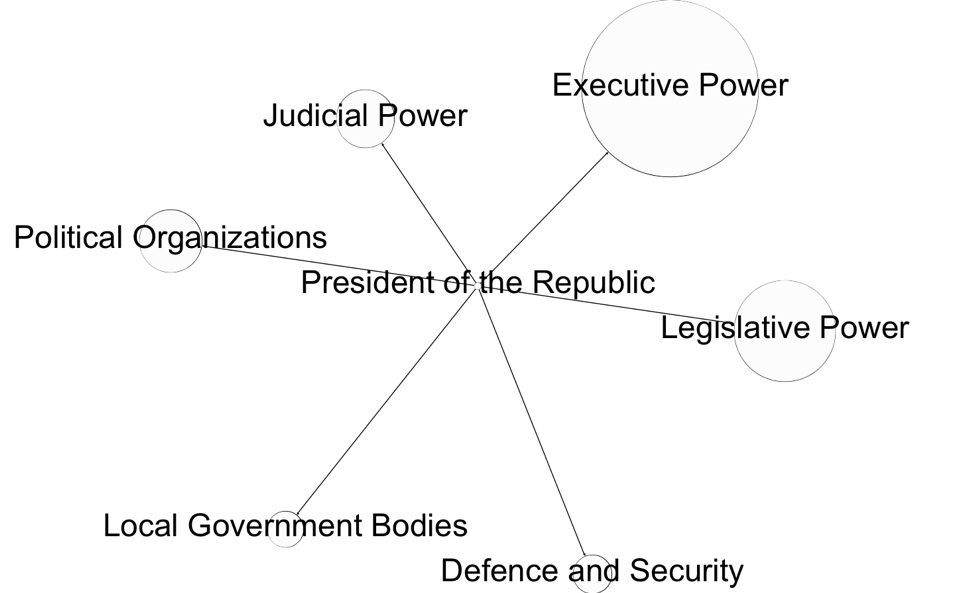}
	\caption{Network of the meetings with the main institutional actors according to the different powers in the Italian political system.}
	\label{fig:netw_powers}
	\vspace{-15pt}
\end{figure}

Suppose we are interested in measuring the relations between the PoR and the majority and minority parliamentary groups in the sphere of internal politics, aggregating the information according to the time frame of each government in the seven-year term of President Giorgio Napolitano. In this case, PC-ACE allows us to perform our analysis in two ways. The first option is to employ a graphical interface which uses a tree graph to represent the entities of our analysis and their hierarchical relations. PC-ACE allows us to represent the structure of the imported data as shown in Figure \ref{fig:qmform}; to select the entries corresponding to the categories shown in the figure (eventually applying some filters to them i.e. selecting all entries which have a particular value in the ``Government'' category) and retrieve them with this interface. 
The second option is to use the Structured Query Language (SQL) and access the data stored in PC-ACE relational database directly. Either way, we can extract data from the hierarchical structure we created and analyze it with the statistical tools available in PC-ACE or using the SQL language.\\
Within PC-ACE, we are also able to create some charts from the data we have imported in the database. For example, as shown in Figure \ref{fig:majorityminority}, we can create an histogram of how the number of meetings with members of the majority and minority parliamentary groups changed over time for each government (Prodi II, Berlusconi IV and Monti) during Napolitano's seven-year term.
Another example of the visualizations that PC-ACE can create with the data we imported is shown in Figure \ref{fig:netw_powers}. The network shows at the highest level of abstraction what are the state powers with which the PoR has interacted the most. Figures \ref{fig:majorityminority} and  \ref{fig:netw_powers} moreover, show the actual power of this approach, which is the ability to generalize arbitrarily the detail of the analysis according to each level of the hierarchy of our structure.

\begin{figure}[htb!]
	\includegraphics[scale=0.18]{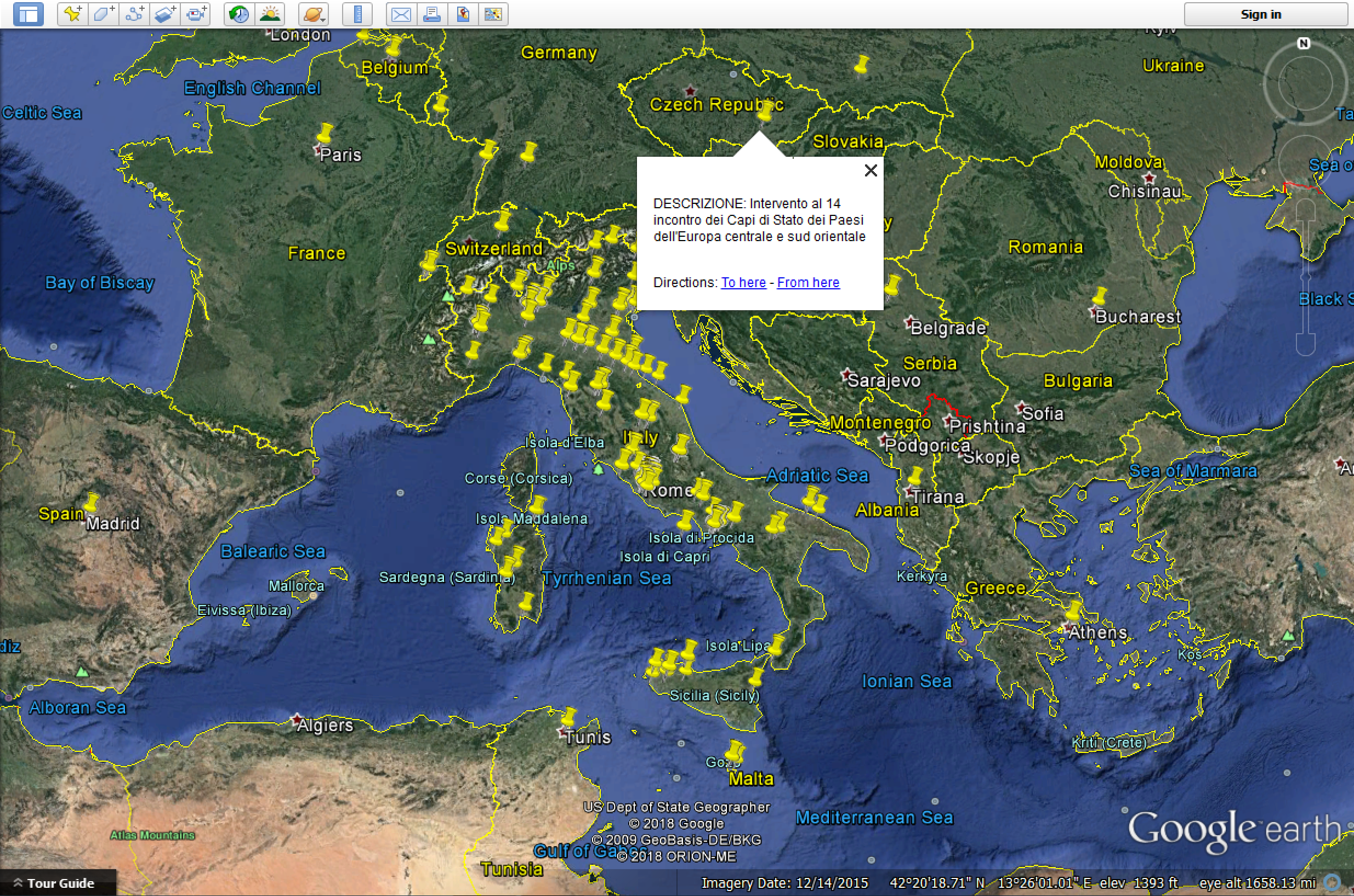}
	\caption{Google Earth Pro visualization of the places visited by President Giorgio Napolitano during his first presidential term.}
	\label{fig:gep}
	\vspace{-15pt}
\end{figure}

Another option that the user has in PC-ACE is to visualize on Google Earth Pro~\footnote{\url{https://www.google.com/earth/}.} the data which has been imported in the software. Since the PoR's Diary contained the location where each meeting was held, we were able to obtain automatically the latitude and longitude coordinates of each location using either Google Earth Pro or other free software such as QGIS~\footnote{\url{http://qgis.org}.} and create automatically the map shown in Figure \ref{fig:gep}. PC-ACE also allows the user to add a description to each point as shown in the Figure and insert any content in it. For instance in our case we added the field "Description" from the Diary.

\section{Conclusion and Future Challenges}
\label{sec:cfc}
In this work, we propose a novel approach to combine the theory of QNA, with NLP tools for the analysis of textual data in our sample case study. We show how to modify the well regulated structure of traditional QNA  -- strongly relying on the invariant narrative structure of documents -- to a more flexible hierarchical structure. Therefore, we opened the way to a more widespread application of this technique for the quantitative analysis of topics which might benefit from its characteristics. 
The main challenge for the application of this technique to the human sciences is the initial investment needed to develop an ad-hoc algorithm for the extraction of information from a general data source. This calls for a collaboration between researchers from different fields, as shown in this sample case study. In fact, as for now, there cannot be one solution which satisfies the needs of all researchers.
However, the ``democratization'' of machine learning approaches and NLP techniques -- such as NER or topic modeling -- is making these approaches more accessible to the general public. They have also become easier to employ in a modular way, as elements of an information extraction and categorization pipeline, that can be customized each time to the user's needs.
Finally, the evolution of software for data analysis such as PC-ACE, fills the gap of competences of potential users who can be guided in the process of data analysis and visualization.

\bibliographystyle{ACM-Reference-Format}
\bibliography{main}

\end{document}